\pdfoutput=1
\documentclass[journal]{IEEEtran}
\usepackage{cite}
\usepackage{amsmath,amssymb,amsfonts}
\usepackage{algorithmic}
\usepackage{graphicx}
\usepackage{textcomp}
\usepackage[nolist]{acronym}
\usepackage{xspace}
\usepackage[capitalise]{cleveref}
\Crefname{figure}{Fig.}{Figs.}
\usepackage{subcaption}
\usepackage{ifthen}
\usepackage{booktabs,tabularx}
\newcolumntype{C}{>{\centering\arraybackslash}X}
\usepackage{xcolor}
\usepackage{tikz}
\usetikzlibrary{positioning,calc,shapes.geometric}
\usepackage{bm}
\usepackage{marvosym}
\usepackage{numprint}
\usepackage{siunitx}
\usepackage{dirtytalk}
\def\etal{\emph{et al.}\xspace}
\def\eg{e.\,g.\xspace}
\def\ie{i.\,e.\xspace}
\def\unet{U-net}

\def\msmf{MS-MF}
\def\mscct{MS-CCT}
\def\mvmr{MV-MR}

\def\mfA{Aperio ScanScope CS2}
\def\mfB{NanoZoomer 2.0-HT}
\def\mfC{NanoZoomer S360}
\def\mfD{Pannoramic 250 Flash III}
\def\mfE{Pannoramic SCAN II}
\def\mfF{SG60}

\def\sccA{Aperio ScanScope CS2}
\def\sccB{NanoZoomer S210}
\def\sccC{NanoZoomer 2.0-HT}
\def\sccD{Pannoramic 1000}
\def\sccE{Aperio GT450}

\def\mrA{MAGNETOM Avanto}
\def\mrB{Achieva}
\def\mrC{Signa Excite}
\def\mrD{Vantage Orian}

\begin{document}
\title{Rethinking U-net Skip Connections for Biomedical Image Segmentation}
\author{Frauke Wilm, Jonas Ammeling, Mathias Öttl, Rutger~H.J. Fick,  Marc Aubreville, Katharina Breininger
\thanks{This work was partially funded by the German Research Foundation (DFG) project 460333672 CRC1540 EBM. F. Wilm acknowledges financial support received by Merck Healthcare KGaA and the scientific exchange with T. Mrowiec during method development. K. Breininger acknowledges support by d.hip campus - Bavarian aim in form of a faculty endowment. J. Ammeling and M. Aubreville acknowledge support from the Bavarian Institute of Digital Transformation (Project ReGInA).}
\thanks{F. Wilm and M. Öttl are with the Department Artificial Intelligence in Biomedical Engineering and with the Pattern Recognition Lab, Friedrich-Alexander-Universität (FAU) Erlangen-Nürnberg, Erlangen, Germany (email: frauke.wilm@fau.de, mathias.oettl@fau.de).}
\thanks{J. Ammeling and M. Aubreville are with the Technische Hochschule Ingolstadt, Germany (email: jonas.ammeling@thi.de, marc.aubreville@thi.de).}
\thanks{R. Fick is with Tribun Health, Paris, France (email: rfick@tribun.health).}
\thanks{K. Breininger is with the Department Artificial Intelligence in Biomedical Engineering, FAU Erlangen-Nürnberg, Erlangen, Germany (email: katharina.breininger@fau.de).} 
\thanks{M. Aubreville and K. Breininger are joint senior authors of this work.}
}

\markboth{submitted to IEEE Transactions of Medical Imaging}%
{Wilm \MakeLowercase{\textit{et al.}}: Rethinking U-net Skip Connections for Biomedical Image Segmentation}
\maketitle

\begin{abstract}
The \unet{} architecture has significantly impacted deep learning-based segmentation of medical images. Through the integration of long-range skip connections, it facilitated the preservation of high-resolution features. Out-of-distribution data can, however, substantially impede the performance of neural networks. Previous works showed that the trained network layers differ in their susceptibility to this domain shift, \eg, shallow layers are more affected than deeper layers. In this work, we investigate the implications of this observation of layer sensitivity to domain shifts of \unet-style segmentation networks. By copying features of shallow layers to corresponding decoder blocks, these bear the risk of re-introducing domain-specific information. We used a synthetic dataset to model different levels of data distribution shifts and evaluated the impact on downstream segmentation performance. We quantified the inherent domain susceptibility of each network layer, using the Hellinger distance. These experiments confirmed the higher domain susceptibility of earlier network layers. When gradually removing skip connections, a decrease in domain susceptibility of deeper layers could be observed. For downstream segmentation performance, the original \unet{} outperformed the variant without any skip connections. The best performance, however, was achieved when removing the uppermost skip connection---not only in the presence of domain shifts but also for in-domain test data. We validated our results on three clinical datasets---two histopathology datasets and one magnetic resonance dataset---with performance increases of up to 10\,\% in-domain and 13\,\% cross-domain when removing the uppermost skip connection.    
\end{abstract}


\begin{figure}[t]
\centering
\resizebox{\linewidth}{!}{
\begin{tikzpicture}
        % LDM Blocks
        \node (rect-l0) at (0,1.5) [draw, minimum width=0.25cm,minimum height=2.5cm, anchor=south west] {};
        \node (rect-r0) at (9.25,1.5) [draw, minimum width=0.25cm,minimum height=2.5cm, anchor=south west] {};
        \node (rect-l1) at (1,0.5) [draw, minimum width=0.5cm,minimum height=2cm, anchor=south west] {};
        \node (rect-r1) at (8,0.5) [draw,minimum width=0.5cm,minimum height=2cm, anchor=south west] {};
        \node (rect-l2) at (2,-0.5) [draw,minimum width=0.75cm,minimum height=1.5cm, anchor=south west] {};
        \node (rect-r2) at (6.75,-0.5) [draw,minimum width=0.75cm,minimum height=1.5cm, anchor=south west] {};
        \node (rect-l3) at (3,-1.5) [draw,minimum width=1cm,minimum height=1cm, anchor=south west] {};
        \node (rect-r3) at (5.5,-1.5) [draw,minimum width=1cm,minimum height=1cm, anchor=south west] {};
        \node (rect-m) at (4,-2.5) [draw,minimum width=1.5cm,minimum height=0.5cm, anchor=south west] {};

        % Skip Connections
        \draw[shorten >=0.25cm,shorten <=0.25cm,->](rect-l0.east) -- (rect-r0.west) {};
        \draw[shorten >=0.25cm,shorten <=0.25cm,->](rect-l1.east) -- (rect-r1.west) {};
        \draw[shorten >=0.25cm,shorten <=0.25cm,->](rect-l2.east) -- (rect-r2.west) {};
        \draw[shorten >=0.25cm,shorten <=0.25cm,->](rect-l3.east) -- (rect-r3.west) {};

        % Pruned
        \node[scale=2, text=teal] (scissors1) at ($(rect-l3)!0.5!(rect-r3)$) {\Cutright};
        \node[scale=2, text=teal] (scissors2) at ($(rect-l2)!0.5!(rect-r2)$) {\Cutright};
        \node[scale=2, text=teal] (scissors3) at ($(rect-l1)!0.5!(rect-r1)$) {\Cutright};
        \node[scale=2, text=teal, label={[scale=0.75, text=teal, align=center]L4- \\ pruned}] (scissors4) at ($(rect-l0)!0.5!(rect-r0)$) {\Cutright};

        \node[scale=2, text=purple] (scissors5) at ($(scissors2)-(1, 0)$) {\Cutright};
        \node[scale=2, text=purple] (scissors6) at ($(scissors3)-(1, 0)$) {\Cutright};
        \node[scale=2, text=purple, label={[scale=0.75, text=purple, align=center]L3- \\ pruned}] (scissors7) at ($(scissors4)-(1, 0)$) {\Cutright};

        \node[scale=2, text=orange] (scissors8) at ($(scissors3)-(2, 0)$) {\Cutright};
        \node[scale=2, text=orange, label={[scale=0.75, text=orange, align=center]L2- \\ pruned}] (scissors9) at ($(scissors4)-(2, 0)$) {\Cutright};

        \node[scale=2, text=cyan, label={[scale=0.75, text=cyan, align=center]L1- \\ pruned}] (scissors10) at ($(scissors4)-(3, 0)$) {\Cutright};
\end{tikzpicture}}
\caption{Schematic illustrations of the baseline \unet{} and the pruned architectures. For the L1-pruned \unet{} (cyan), the uppermost skip connection was removed. Consecutively removing the layer-wise skip connections resulted in the L2-pruned (orange), L3-pruned (red), and L4-pruned (teal) \unet.}
\label{fig:architectures}
\end{figure}
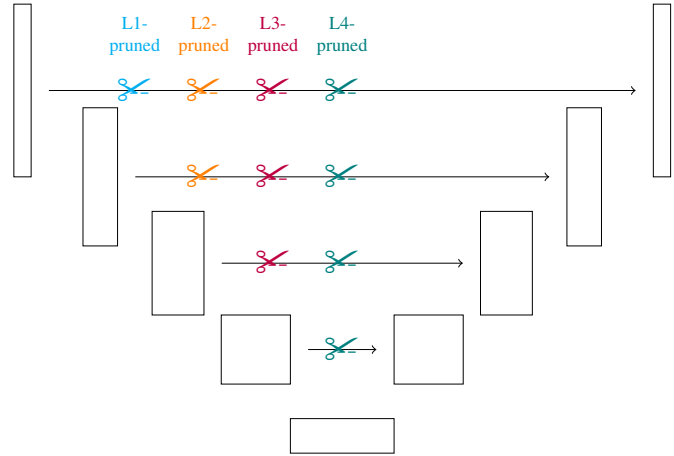

\section{Introduction}
\label{sec:introduction}
\IEEEPARstart{C}{onvolutional} \acp{cnn} have achieved outstanding performance in solving a wide range of medical image analysis tasks, often reaching the performance of trained experts~\cite{rajpurkar2017chexnet, aubreville2020deep}. In 2015, Ronneberger \etal~\cite{ronneberger2015u} presented the \unet{} architecture---a milestone in the field of medical image segmentation. By introducing long-range skip connections that concatenate the feature maps at multiple encoder levels to the corresponding decoder level, the authors demonstrated a better preservation of feature resolutions. Meanwhile, a wide range of architectural modifications has been proposed, and variants of the \unet{} architecture consistently outperform other architectures at public computer vision challenges on medical image segmentation~\cite{isensee2021nnu}. However, the performance of \acp{cnn}---classification and segmentation architectures alike---heavily relies on the training data distribution. Distribution shifts between training and test data, commonly referred to as \emph{domain shift}, can affect \ac{cnn} performance substantially~\cite{quinonero2008dataset}. Opposed to the common belief that early \ac{cnn} layers extract more domain-agnostic features than deeper layers~\cite{yosinski2014transferable}, recent studies on the domain susceptibility of \ac{cnn} architectures~\cite{aljundi2016lightweight,shirokikh2020first} suggest the contrary. These studies reported a higher domain shift susceptibility of earlier network layers both for classification~\cite{aljundi2016lightweight} as well as segmentation~\cite{shirokikh2020first} tasks. We consider these observations particularly relevant for segmentation architectures with skip connections which copy the feature maps from different encoder levels to the corresponding decoder levels. Thereby, they might bear the risk of passing domain-specific features to deeper layers of the network and making predictions less domain-agnostic. Existing works, however, have not explicitly investigated the role of skip connections in this regard.

In this work, we extend previous empirical experiments by explicitly quantifying the domain shift of encoder-decoder \ac{cnn} architectures at different layers. We evaluate the influence on segmentation performance and robustness against domain shifts. In particular, we study the role of skip connections by performing a layer-wise pruning of skip connections, as illustrated in \cref{fig:architectures}. To quantify the inherent domain shift at each \ac{cnn} layer, we utilized the Hellinger distance. Compared to other previously proposed metrics, \eg, representation shift~\cite{stacke2020measuring}, the Hellinger distance can be defined as a bound metric and is scale-invariant. Thereby, it allows for comparing different \ac{cnn} layers with varying dimensionalities and feature scales.
Our contributions can be summarized as follows:
\begin{itemize}
\item modeling of various common domain shifts on a custom synthetic dataset   
\item layer-wise pruning of skip connections and evaluation of downstream segmentation performance
\item utilization of Hellinger distance to quantify the inherent domain shift at each layer of the \ac{cnn}
\item validation of findings on three medical datasets (two histopathology datasets and one cardiac \ac{mr} dataset)
\end {itemize}

We observe a beneficial impact on performance and robustness when removing the upper-most skip connection (L1-pruned), \ie, the L1-pruned model achieved the best in-domain and cross-domain segmentation performance for all datasets. Furthermore, we show that earlier layers are considerably more susceptible to domain shifts than deeper layers, which is in line with the observations of related studies. Based on these results we propose to rethink the use of skip connections in biomedical image segmentation and consider potential skip connection pruning during model development.   
 
\section{Related work}
\subsection{\unet{} for biomedical image segmentation}
Since the original implementation of the vanilla \unet{} in 2015~\cite{ronneberger2015u}, many architectural variants have been proposed. The dense \unet~\cite{cai2020dense} employed dense blocks at all encoder levels to facilitate better reuse of features. The idea of adding dense connections was also pursued by the \unet++ architecture~\cite{zhou2018unet++}, which employed a network of skip connections to propagate the features of one encoder level to multiple decoder levels. The inception \unet~\cite{szegedy2015going} used wider encoder blocks with parallel paths of varying filter kernel sizes to accommodate for features of varying semantic scales, whereas the residual \unet~\cite{he2016deep} used residual blocks in the encoder to allow for deeper blocks at each level. The recurrent-residual (R2) \unet~\cite{alom2019recurrent} additionally utilized recurrent connections for feature accumulation, whereas the attention \unet~\cite{oktay2018attention} added attention blocks and the SE \unet~\cite{roy2018concurrent} squeeze-and-excitation blocks to the vanilla architecture to allow feature re-weighting along the network layers. While all of the abovementioned architectural variants changed the structural design of the encoder blocks or skip connections, all of them retained the original idea of using long-range skip connections to concatenate the encoder output to one or more decoder levels. Even though many works showed that skip connections improve the overall segmentation performance compared to a simple encoder-decoder architecture, recent work by Wang \etal~\cite{wang2022uctransnet} investigated the contribution of each skip connection and found that in some cases individual skip connections can in fact harm the overall performance. Consequently, the authors proposed a transformer-based feature fusion as an alternative to the standard copying of encoder features.       

\iffalse
\begin{figure}[t]
    \centering
    \includegraphics[width=\linewidth]{unets.pdf}
    \caption{Figure taken from Kugelman \etal~\cite{kugelman2022comparison} licensed under CC~BY~4.0 (https://www.creativecommons.org/licenses/by/4.0/). For improved visualization, each architecture is only depicted with two layers. Dashed arrows visualize skip connections.}
    \label{fig:unets}
\end{figure}
\fi

\subsection{Domain susceptibility of \ac{cnn} architectures} 
Early works on transfer learning commonly used fine-tuning of individual network layers for adapting a trained model to a new target domain. These works often re-trained the last \ac{cnn} layers, based on the assumption that early layers extract features that are more generic across domains, \ie, edges or texture descriptors, whereas deeper layers extract more domain-specific features~\cite{yosinski2014transferable}. Recent works studied the sensitivity to domain shifts of individual \ac{cnn} layers in more detail and reported contradictory results~\cite{aljundi2016lightweight,shirokikh2020first}. Aljundi \etal~\cite{aljundi2016lightweight} used the H-divergence~\cite{ben2006analysis} to quantify the domain susceptibility of different filter kernels at different levels and reported a higher domain sensitivity for earlier compared to later layers. Shirokikh \etal~\cite{shirokikh2020first} showed that fine-tuning the early layers of a \unet{} architecture with target domain samples was considerably more effective for domain generalization than fine-tuning deeper layers or even the complete \ac{cnn} when a limited amount of target domain data were available.   

\subsection{Measuring domain shift} 
Many works on domain shift quantification stem from the field of uncertainty quantification and \ac{ood} detection, where the softmax output of the final prediction layer is interpreted as model confidence~\cite{stacke2020measuring}. Recent literature, however, suggests that covariate shifts can already be detected at intermediate network layers, and various metrics to quantify the inherent domain shifts have been proposed~\cite{stacke2020measuring,schilling2021quantifying}. The authors of these works have demonstrated a correlation of these metrics with the model's accuracy and their eligibility to predict the expected performance degradation. Stacke \etal~\cite{stacke2020measuring} proposed the \emph{representation shift}, computed from the Wasserstein distance between the domain-specific feature distributions over all filters in a network layer. Even though the authors demonstrated a high correlation with a decrease in classification accuracy, the representation shift has one important shortcoming: Through the computation from a network's activation maps, the metric becomes unbound and is highly dependent on feature scaling and dimensionality, which prevents comparison across models or datasets. This shortcoming can be alleviated by distribution distance measures such as the Hellinger distance. Compared to other f-divergences, \eg, the \ac{kl} Divergence, the Hellinger distance is symmetric, bound between \num{0} and \num{1}, and fulfills the triangle inequality. These characteristics allow comparing the Hellinger distance across layers with a different number of neurons. Previous work~\cite{gonzalez2013class} has demonstrated the suitability of the Hellinger distance for quantifying distribution shifts between training and test data as well as high robustness against the base \ac{cnn} performance.

\section{Materials}
To reduce the influence of label noise and potential hidden domain shifts, \eg batch effects introduced during image acquisition, we first created a synthetic dataset to model different domain shift scenarios explicitly and study their influence on the model's predictive performance. We modeled controlled changes in brightness, saturation, and contrast. In histopathology, changes in brightness and saturation are common between images acquired in different environments~\cite{aubreville2023mitosis}, whereas changes in contrast between devices are common for many imaging modalities. The observations made on the synthetic dataset were then validated on three medical datasets---two multi-scanner histopathology datasets for mitotic figure detection~\cite{aubreville2020completely} and cutaneous tumor segmentation~\cite{wilm2023multi}, and a publicly available multi-center, multi-vendor, multi-disease \ac{mr} dataset for cardiac segmentation. The clinical datasets were gathered during previous publications, and appropriate institutional review board approval was secured for those publications. For more information, we refer to the relevant manuscripts.

\subsection{Synthetic malaria dataset}
\begin{figure}[t]
    \centering
    \begin{minipage}[b]{0.405\linewidth}
        \centering
        \includegraphics[height=\linewidth, angle=90]{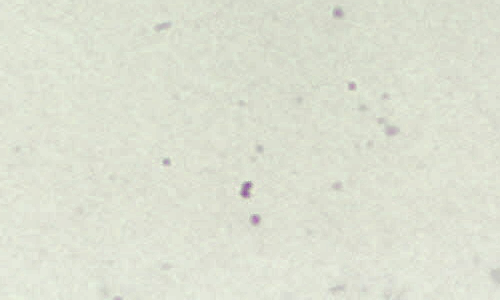}
    \end{minipage}
    \hfill
    \begin{minipage}[b]{0.27\linewidth}
        \centering
        \includegraphics[width=\linewidth]{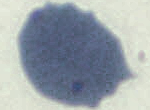}\par
        \vspace{1em}
        \includegraphics[width=\linewidth]{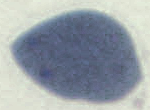}\par
        \vspace{1em}
        \includegraphics[width=\linewidth]{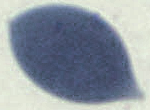}
    \end{minipage}
    \hfill
    \begin{minipage}[b]{0.27\linewidth}
        \centering
        \includegraphics[width=\linewidth]{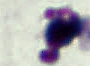}\par
        \vspace{1em}
        \includegraphics[width=\linewidth]{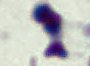}\par
        \vspace{1em}
        \includegraphics[width=\linewidth]{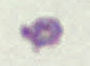}
    \end{minipage}
    \caption{Templates used for the creation of the synthetic malaria dataset: background (left column), cells (middle column), and artifacts (right column).}
    \label{fig:synthetic}
\end{figure}

\begin{figure}[t]
\centering
\includegraphics[width=0.25\linewidth]{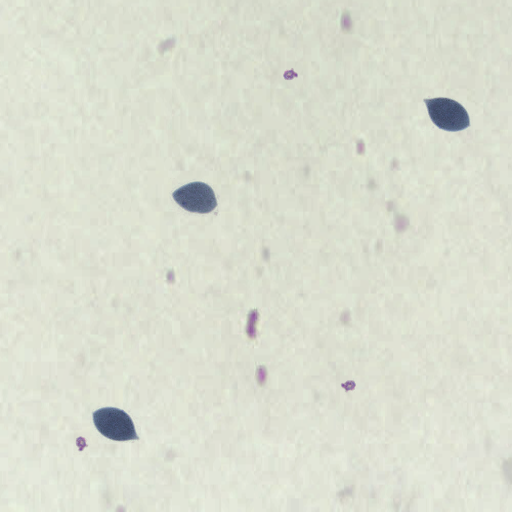}%
\includegraphics[width=0.25\linewidth]{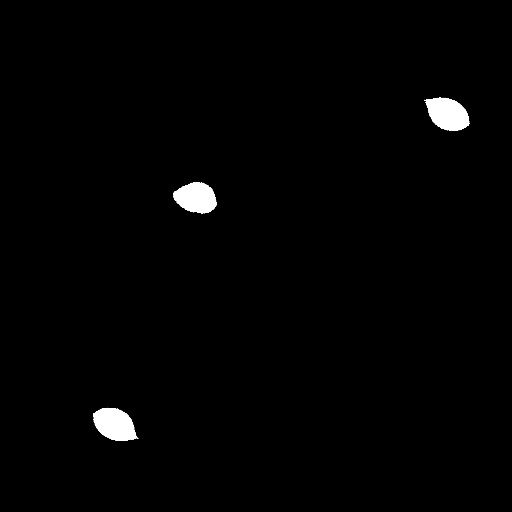}%
\includegraphics[width=0.25\linewidth]{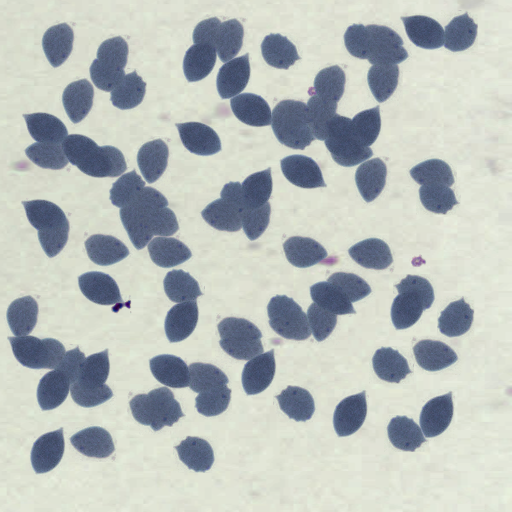}%
\includegraphics[width=0.25\linewidth]{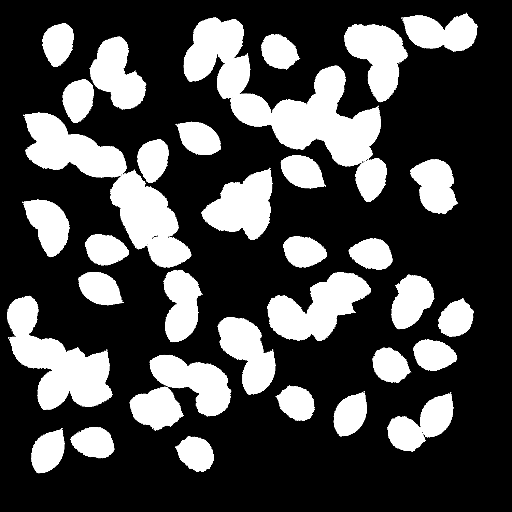}
\caption{Exemplary images and corresponding masks of the synthetically created malaria dataset.}
\label{fig:result}
\end{figure}

The synthetic malaria dataset was created based on the BBBC041v1 image set, available from the Broad Bioimage Benchmark Collection~\cite{ljosa2012annotated}. The dataset creation followed a description on kaggle\footnote{https://www.kaggle.com/code/vbookshelf/how-to-generate-artificial-cell-images/notebook}: First, a random sample image was selected from the dataset. From this sample, one background patch, three malaria cell patches, and three artifacts were cut out (visualized in \cref{fig:synthetic}). From these templates, \num{1000} synthetic images were created. For each synthetic image, the following image generation process was followed:
\begin{enumerate}
    \item randomly rotate the background template by $\theta$ degrees and resize to \num{1500}\,$\times$\,\num{1500} pixels, $\theta~\in~\{\numlist{0;90;180;270}\}$;
    \item randomly sample $n$ malaria cells from the three templates and place them at random (x,y)-positions, $n \in [\numlist{0;100}]$; 
    \item randomly sample three artifacts from the three templates and place them at random (x,y)-positions;
    \item resize the image to \num{512}\,$\times$\,\num{512} pixels.
\end{enumerate}
The ground truth masks were created by thresholding the cell templates and placing the cell masks in an empty mask at the same (x,y)-positions of the corresponding synthetic image. By sampling the number of malaria cells in the interval of $1 \leq n \leq 100$, samples with varying cell densities were created. \Cref{fig:result} visualizes an exemplary low-density and high-density image, created with the described method. After image generation, \SI{20}{\percent} of the images were defined as a hold-out test set. An equal distribution of low-and high-density images was ensured in the training and test split.    

\subsection{Multi-scanner mitotic figure dataset (\msmf)}
\begin{figure}[t]
\centering
\includegraphics[height=1.25cm]{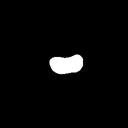}%
\includegraphics[height=1.25cm]{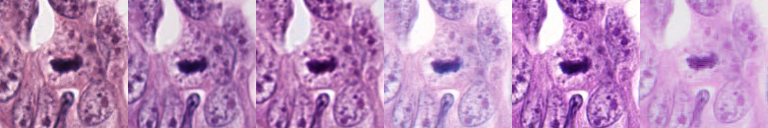} \\ 
\vspace{0.1cm}
\includegraphics[height=1.25cm]{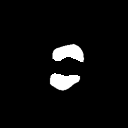}%
\includegraphics[height=1.25cm]{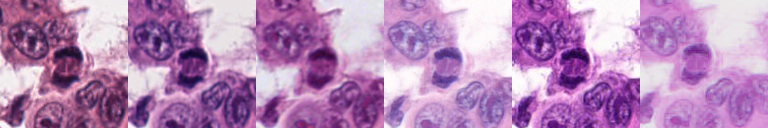} \\ 
\vspace{0.1cm}
\includegraphics[height=1.25cm]{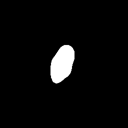}%
\includegraphics[height=1.25cm]{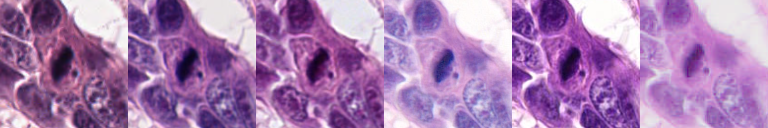} \\ 
\setlength\tabcolsep{2pt}%
\begin{tabularx}{0.98\linewidth}{CCCCCCC}
 & A & B & C & D & E & F \\
\end{tabularx}
\caption{Multi-scanner histopathology dataset for mitotic figure segmentation. Each row visualizes a concentric mitotic figure on the same tissue sample digitized with six scanning systems. A:~\mfA{} (Leica), B:~\mfB{} (Hamamatsu), C:~\mfC{} (Hamamatsu), D:~\mfD{} (3DHISTECH), E:~\mfE{} (3DHISTECH), F:~\mfF{} (Philips).} 
\label{fig:mfs}
\end{figure}

The first medical dataset used for validating the toy dataset results comprised a multi-scanner histopathology dataset focusing on the task of mitotic figure segmentation. The original dataset is composed of \num{21} \ac{cmc} \acp{wsi}, digitized with the \mfA{} scanning system, and was originally published by Aubreville \etal~\cite{aubreville2020completely}. For the experiments presented in this study, 19 samples were re-scanned with five additional systems (two glass slides were damaged during sample shipment). \Cref{tab:vendors} provides an overview of the scanner manufacturers and image resolutions. For \ac{wsi} alignment, we used a quadtree-based registration algorithm by Marzahl \etal~\cite{marzahl2021robust}.  The Aperio \acp{wsi} were annotated for mitotic figures following a two-stage semi-automatic labeling process. This resulted in a total number of \num{14154} point annotations. For this work, we converted these point annotations to segmentation masks using a pre-trained \unet{} for automatic mask generation and then manually corrected all generated masks. Finally, we extracted object-centric patches sized \num{128}\,$\times$\,\num{128} pixels, which were used for training a \ac{cnn} for mitotic figure segmentation. \Cref{fig:mfs} visualizes a mitotic figure from the same tissue sample digitized with six slide scanners of different scanners. For \ac{cnn} training, only the \mfA{} patches were used. Testing was then performed across all scanner domains. From the \num{19} \acp{wsi}, four ($\sim$ \SI{20}{\percent}) were selected as a hold-out test set with a total number of \num{5605} annotations. Due to registration inaccuracies, the final test set only comprised \num{5245} multi-scanner mitotic figure patches. In the following sections, we will refer to the dataset as \msmf{} dataset.   

\subsection{Multi-scanner canine cutaneous tumor dataset (\mscct)}
\begin{figure}[t]
\centering
\includegraphics[height=1.25cm]{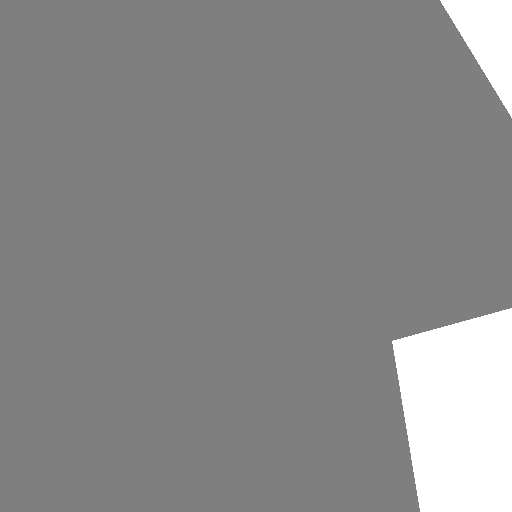}%
\includegraphics[height=1.25cm]{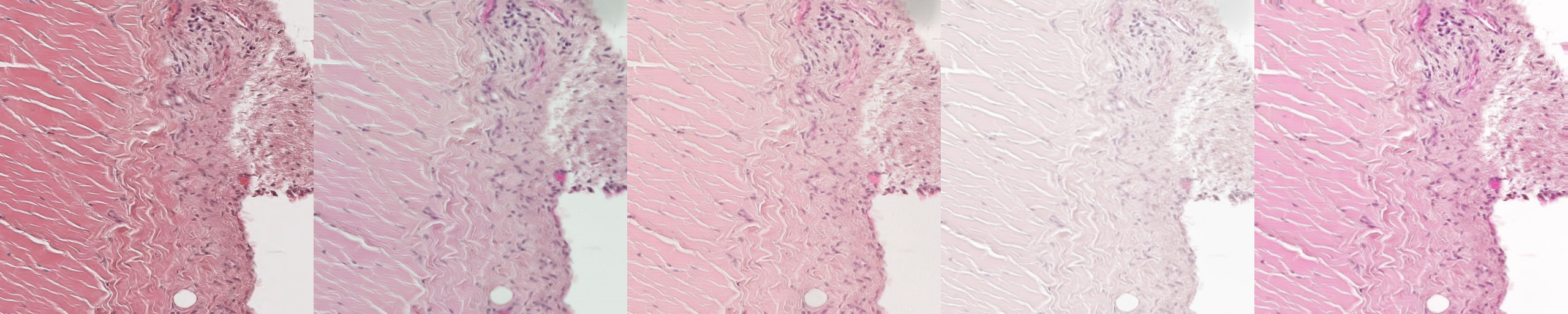} \\ 
\vspace{0.1cm}
\includegraphics[height=1.25cm]{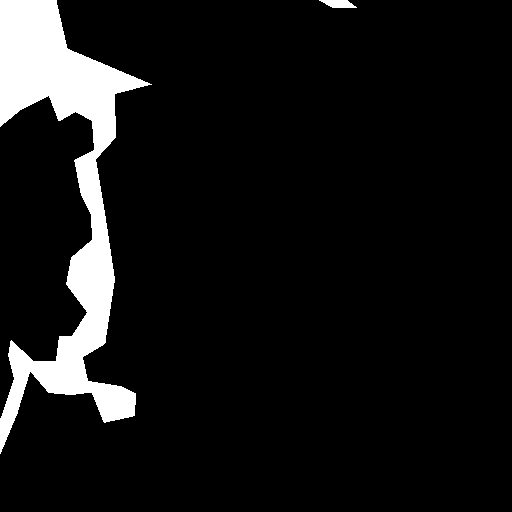}%
\includegraphics[height=1.25cm]{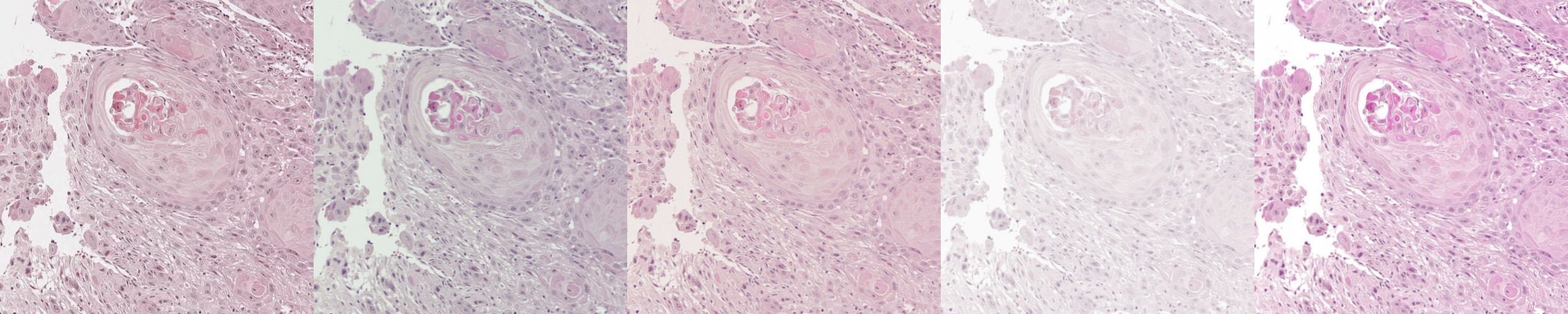} \\ 
\vspace{0.1cm}
\includegraphics[height=1.25cm]{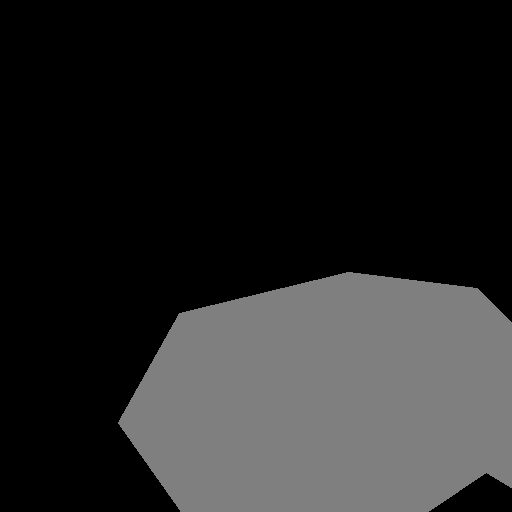}%
\includegraphics[height=1.25cm]{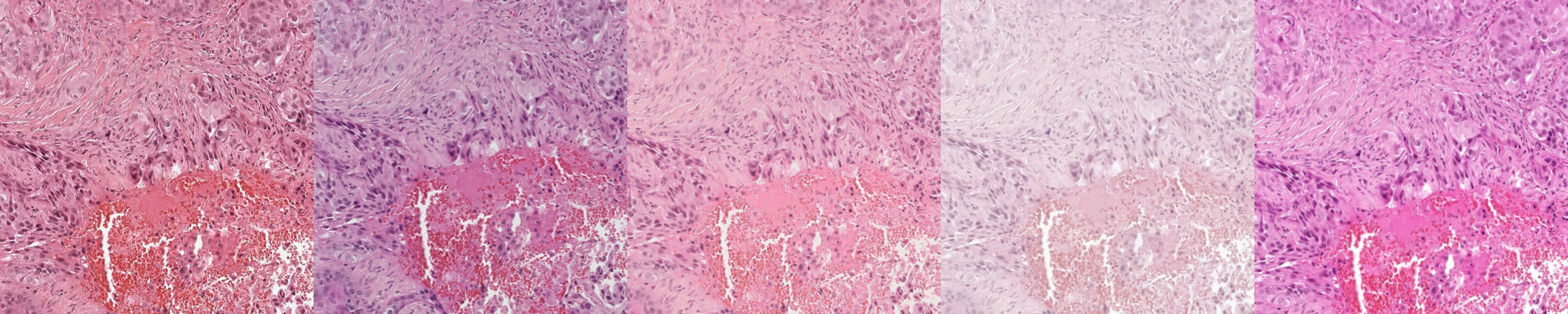} \\ 
\setlength\tabcolsep{2pt}%
\begin{tabularx}{0.85\linewidth}{CCCCCC}
 & A & B & C & D & E \\
\end{tabularx}
\caption{Multi-scanner histopathology dataset for tumor segmentation. Each row visualizes a patch from the same tissue sample digitized with five scanning systems (white: background, gray: non-tumor, black: tumor). A:~\sccA{} (Leica), B:~\sccB{} (Hamamatsu), C:~\sccC{} (Hamamatsu), D:~\sccD{} (3DHISTECH), E:~\sccE{} (Leica).} 
\label{fig:cutaneous}
\end{figure}

The second medical dataset also comprised a multi-scanner histopathology dataset. The dataset is a subset of the \ac{catch} dataset~\cite{wilm2022pan}, a collection of originally \num{350} \acp{wsi} of seven canine cutaneous tumor subtypes (\num{50} \acp{wsi} per subtype), digitized with the \sccA{} scanning system. For this work, we used the \ac{scc} subset and re-scanned the samples with five additional systems (details in \cref{tab:vendors}). Due to scanning artifacts in at least one of the scans, six samples were removed from the dataset, resulting in a total of \num{44} samples digitized with five scanning systems. A low-resolution version of the dataset is publicly available on Zenodo\footnote{https://doi.org/10.5281/zenodo.7418555} (due to file size restriction the \acp{wsi} could not be uploaded in full resolution) and a detailed evaluation of the scanner imaging statistics can be found in~\cite{wilm2023multi}. \Cref{fig:cutaneous} illustrates a registered patch from the multi-scanner dataset. The original \ac{catch} dataset provided annotations for \num{13} histologic classes (including the seven tumor subtypes). For this work, we have simplified the annotations and sorted them into a tumor and non-tumor class. Furthermore, we used Otsu thresholding~\cite{otsu1979threshold} to detect the white slide background, resulting in a three-class segmentation task. We randomly selected nine of the samples ($\sim$ \SI{20}{\percent}) as a hold-out test set. To accommodate for class imbalances during \ac{cnn} training, we followed a custom sampling strategy: During each epoch, we sampled \num{32} patches per \ac{wsi}, sized \num{512}\,$\times$\,\num{512} pixels, using pre-defined class weights. We used the same frequency for sampling patches within tumor and non-tumor annotations, and \SI{10}{\percent} for slide background.  During testing, we limited the inference to patches with a tissue content above \SI{50}{\percent}. Thereby, we accounted for variations in the scanned slide background of the scanner vendors, as some manufacturers include automatic tissue detection before scanning, which results in considerably less white slide background in the \ac{wsi}. In total, the final test set constituted \num{4778} multi-scanner patches. In the following sections, we will refer to the dataset as \mscct{} dataset. 

\subsection{Multi-vendor magnetic resonance dataset (\mvmr)}
\begin{figure}[t]
\centering
\includegraphics[width=0.25\linewidth]{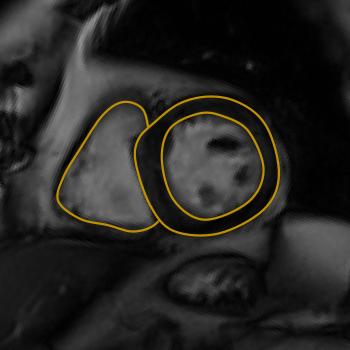}%
\includegraphics[width=0.25\linewidth]{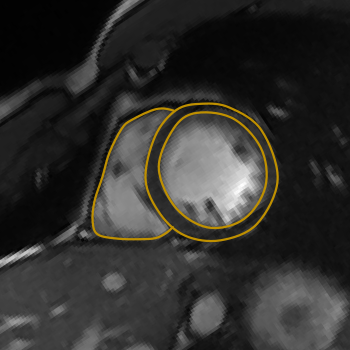}%
\includegraphics[width=0.25\linewidth]{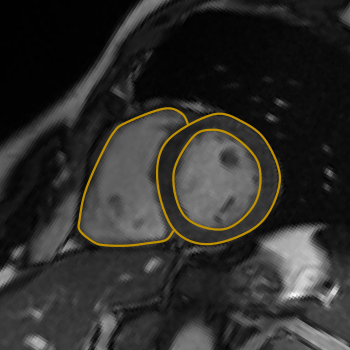}%
\includegraphics[width=0.25\linewidth]{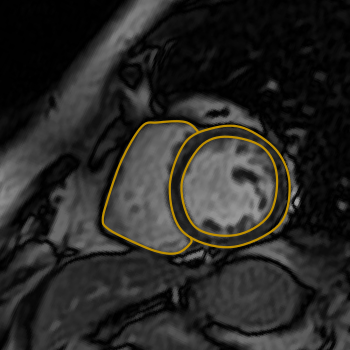}\\
\setlength\tabcolsep{2pt}%
\begin{tabularx}{\linewidth}{CCCC}
 A & B & C & D \\
\end{tabularx}
\caption{Multi-vendor magnetic resonance dataset for cardiac segmentation. Patches visualize four anatomically similar subjects digitized with four different vendors. Figure adapted from Campello \etal~\cite{campello2021multi} under CC BY 4.0 license. A:~\mrA{} (Siemens), B:~\mrB{} (Philips), C:~\mrC{} (General Electric), D:~\mrD{} (Canon).} 
\label{fig:mri}
\end{figure}

Finally, we used a publicly available multi-center, multi-vendor, multi-disease \ac{mr} dataset~\cite{campello2021multi}, published in the context of the M\&Ms challenge for the task of cardiac segmentation. The original dataset comprises \num{375} cardiac \ac{mr} images from four different scanner vendors in six hospitals. We limited our experiments to one hospital per vendor and to samples for which a segmentation was provided. Vendor details can be obtained from \cref{tab:vendors}. While the other datasets provide local correspondences (same slide scanned with different scanners), the samples in this dataset are acquired from different samples. While this may increase differences between domains, \eg, due to site-specific patient demographics, the dataset is still suited for domain shift experiments due to substantial similarities between the images as illustrated in \cref{fig:mri}. For \ac{cnn} training, we used the official challenge train split of the \mrA{} scanner with \num{75} subjects. We limited the test set to an equal number of samples per scanner, resulting in \num{16} subjects per vendor. The provided annotations covered three biological regions: the left and right ventricle cavities (LV and RV, respectively), and the left ventricle myocardium (MYO). For \ac{cnn} training and inference, each projection slice along the temporal and z-axis was considered a separate sample and spatially cropped to a multiple of \num{16} to match the input requirements of the \unet{} architecture. While this resulted in different input sizes for the different scanners (\mrA: \num{192}\,$\times$\,\num{192}, \mrB: \num{320}\,$\times$\,\num{320}, \mrC: \num{256}\,$\times$\,\num{256}, \mrD: \num{416}\,$\times$\,\num{416}), it circumvented the need for re-sizing and potentially interfering with the data-inherent domain shift. For \ac{mr} normalization, we used per-slice re-scaling.

\begin{table}[t]
\caption{Technical specifications of digitization systems used for creating the multi-domain medical datasets.}
\label{tab:vendors}
\setlength{\tabcolsep}{3pt}
    \centering
    \begin{tabular}{p{10pt}p{90pt}p{60pt}p{55pt}}
    %\begin{tabular}{llll}
    \toprule
    \multicolumn{4}{l}{\msmf} \\
    \midrule
        \textbf{A} & \textbf{\mfA} & \textbf{Leica} & \SI{0.25}{\micro\meter\per pixel} \\
        B & \mfB & Hamamatsu & \SI{0.23}{\micro\meter\per pixel}\\
        C & \mfC & Hamamatsu & \SI{0.23}{\micro\meter\per pixel} \\
        D & \mfD & 3DHISTECH & \SI{0.12}{\micro\meter\per pixel} \\
        E & \mfE & 3DHISTECH & \SI{0.25}{\micro\meter\per pixel} \\
        F & \mfF & Philips  & \SI{0.25}{\micro\meter\per pixel} \\
    \midrule
    \multicolumn{4}{l}{\mscct} \\
    \midrule
        \textbf{A} & \textbf{\sccA} & \textbf{Leica} & \SI{0.25}{\micro\meter\per pixel} \\
        B & \sccB & Hamamatsu &  \SI{0.22}{\micro\meter\per pixel} \\
        C & \sccC & Hamamatsu & \SI{0.23}{\micro\meter\per pixel} \\
        D & \sccD & 3DHISTECH & \SI{0.25}{\micro\meter\per pixel} \\
        E & \sccE & Leica & \SI{0.26}{\micro\meter\per pixel} \\
    \midrule
    \multicolumn{4}{l}{\mvmr} \\
    \midrule
        \textbf{A} & \textbf{\mrA} & \textbf{Siemens} & \SI{1.32}{\milli\meter\per pixel} \\
        B & \mrB & Philips & \SI{1.20} {\milli\meter\per pixel} \\
        C & \mrC & General Electric & \SI{1.36} {\milli\meter\per pixel} \\
        D & \mrD & Canon & \SI{0.85} {\milli\meter\per pixel} \\

    \bottomrule
    \end{tabular}
\end{table}

\section{Methods}
To evaluate the influence of skip connections on the robustness of encoder-decoder architectures against domain shifts, we first used the synthetic dataset to model specific domain shift scenarios. We then used the Hellinger distance to measure the inherent domain shift in the feature embeddings of the trained \unet{}. Afterward, we conducted various experiments on removing particular skip connections and evaluated the influence on segmentation performance and domain susceptibility of the architecture. We detail our approaches in the following paragraphs.  

\subsection{Domain shift modeling}
\label{subsec:domain-shift}
We used the synthetic malaria dataset to explicitly model common domain shift scenarios, namely variation of brightness, contrast, and saturation. For modeling brightness, we scaled the input image $\bm{x}$ with a factor $\lambda_b$:
\begin{equation}
    \bm{x}^* = \lambda_b \cdot \bm{x}, \enspace  0.5 \leq \lambda_b \leq 1.75 \enspace .
\end{equation}%
For contrast modeling, we re-adjusted the contrast using a factor $\lambda_c$ and a \ac{lut}, computed from the mean of all $M$ pixels of the grayscale-converted image $g(\bm{x})$:
\begin{equation}
\begin{split}
\overline{\bm{x}} &= \frac{1}{M} \sum_{j=1}^M g(\bm{x})_j \\
\text{lut} &= [0,...,255]^\top \cdot \lambda_c, \enspace  0.5 \leq \lambda_c \leq 1.75 \\
\text{lut} &= \text{lut} + \overline{\bm{x}} \cdot (1 - \lambda_c) \\
\text{lut} &= \max(0, \min(\text{lut}, 255)) \\
 x^*_j & = \text{lut}[x_j], \enspace \forall j \in [0, \dots, M] \enspace .
\end{split}
\end{equation}%
Finally, we modeled the saturation, using a factor $\lambda_s$:
\begin{equation}
    \bm{x}^* = \lambda_s \cdot \bm{x} + (1-\lambda_s) \cdot g(\bm{x}), \enspace  0 \leq \lambda_s \leq 0.75 \enspace .
\end{equation}

For the medical segmentation datasets, domain shifts were inherently provided by the dataset, \ie, different visual appearances caused by different slide scanners (\msmf, \mscct) or \ac{mr} vendors (\mvmr).

\subsection{Hellinger distance for domain shift quantification}
\label{subsec:hellinger}
After model training, we evaluated the inherent domain shift at various network layers using the Hellinger distance. Let $X=\{\bm{x}_1,\cdots,\bm{x}_N\}$ denote the set of original test images and $X^*=\{\bm{x}^*_1,\cdots,\bm{x}^*_N\}$ the set of augmented images generated by a fixed augmentation setting (\eg, $ \bm{x}^* = \lambda_b \cdot \bm{x}|\lambda_b = 0.75$) or a \say{natural} domain shift, \eg, scanner-induced. Furthermore, let $\phi(\bm{x}_i)^{(l)}$ denote the feature map of $\bm{x}_i$ at layer $l, \forall i \in [1, \dots, N]$. To save computational resources, each feature map was average-pooled with an adjusted stride to yield a uniform spatial size of $\frac{p_x}{32}\,\times\,\frac{p_x}{32}$. Here, $p_x$ denotes the side length of the image $\bm{x}$. Afterward, the feature maps were flattened along the spatial dimensions, yielding a set of $K$ feature vectors $\left(\phi(\bm{x}_i)_1^{(l)},\cdots,\phi(\bm{x}_i)_K^{(l)}\right)$, with $K=\left(\frac{p_x}{32}\right)^2$ and $\phi(\bm x_i)_k^{(l)} \in \mathbb R^D$. 

We computed the Hellinger distance for each dimension $d \in D$ and then averaged across all feature channels. For efficient computation of the Hellinger distance, we calculated two binned relative frequency distributions $P_{d}^{(l)}(X)$ and $Q_{d}^{(l)}(X^*)$ from the two image sets $X$ and $X^*$. For each bin $b \in [0,\dots,B-1]$ the entry $P_{d,b}^{(l)}(X)$ of the binned relative distributions can be determined as:

\begin{equation}
P_{d,b}^{(l)}(X)  = \frac{1}{KN} \sum_i \sum_k \delta \left(\phi(\bm x_i )_{d,k}^{(l)}, \frac{b}{B}, \frac{b+1}{B} \right) \enspace ,
\end{equation}%
with $\delta{}$ defining the indicator function of a bin with relative range $z = \left[ z_1, z_2 \right]$ as:

\begin{equation}
\delta(x, z) = \begin{cases} 1 & \mathrm{if} ~  z_1 \alpha_d^{(l)} + \beta_d^{(l)} \leq x < z_2\alpha_d^{(l)} + \beta_d^{(l)} \\ 0  & \mathrm{else}\end{cases} 
\end{equation}%
and the range being scaled in the joint value range of the features as:

\begin{eqnarray}
\alpha_d^{(l)} &=& \max_{i,k} \left\{\phi(\bm x_i )_{d,k}^{(l)}, \phi(\bm x_i^* )_{d,k}^{(l)} \right\}  \nonumber \\ && - \min_{i,k} \left\{\phi(\bm x_i )_{d,k}^{(l)}, \phi(\bm x_i^* )_{d,k}^{(l)} \right\} \\
\beta_d^{(l)} &=& \min_{i,k} \left\{\phi(\bm x_i )_{d,k}^{(l)}, \phi(\bm x_i^* )_{d,k}^{(l)} \right\}.
\end{eqnarray}%
Likewise, we define the binned relative distribution $Q_{d}^{(l)}(x^*)$ for image set $X^*$. The Bhattacharyya coefficient, which models the distribution overlap within the dimension $d$, could then be estimated as
\begin{equation}
BC(P_d^{(l)}(X),Q_d^{(l)}(X^*)) = \sum_{b=0}^{B-1}\sqrt{P_{d,b}^{(l)}(X) \cdot Q_{d,b}^{(l)}(X^*)} \enspace . 
\end{equation}%
From the Bhattacharyya coefficient, the Hellinger distance at dimension $d$ and layer $l$ can be computed as
\begin{equation}
\Delta(P_d^{(l)}(X),Q_d^{(l)}(X^*)) = \sqrt{1-BC(P_d^{(l)}(X),Q_d^{(l)}(X^*))} \enspace . 
\end{equation}%
For the computation of the domain shift at layer $l$ with feature dimensionality $D$, we averaged the Hellinger distance across all feature dimensions $d \in D^{(l)}$
\begin{equation} 
\Delta^{(l)} = \frac{1}{D^{(l)}} \sum_{d=1}^{D^{(l)}}\Delta(P_d^{(l)}(X),Q_d^{(l)}(X^*)) \enspace .
\end{equation}

\subsection{Skip connection pruning}
We first trained a baseline residual \unet{} with a ResNet~\cite{he2016deep} encoder and skip connections between encoder and decoder blocks. We experimented with two encoder sizes---ResNet18 and ResNet34, \ie, the configurations of the ResNet architecture that utilize 18 and 34 layers, respectively. We then pruned the skip connections from top to bottom, resulting in an L1-pruned, L2-pruned, L3-pruned, and L4-pruned \unet{}. \Cref{fig:architectures} schematically illustrates the baseline and the pruned architecture designs. \Cref{tab:unets} summarizes the number of trainable parameters for both encoder sizes and demonstrates that the parameter count is only mildly affected by skip connection pruning. 

For each model, we repeated the training five times using a stratified split of the training set and used the validation fold for model selection. For all models, we used an encoder pre-trained on ImageNet~\cite{russakovsky2015}. The model was trained with a batch size of \num{8}, the Adam optimizer, and a learning rate step decay schedule with an initial learning rate of \num{e-4}. The model was optimized with a combined cross-entropy and Dice~\cite{sudre2017generalised} loss and trained for \num{50} epochs by which we observed model convergence. We used the validation split for model selection, based on the highest \ac{miou}. We used standard affine transformations, \ie flipping, cropping, rotation, and scaling, but refrained from using color augmentation or normalization to better highlight the impact of visual domain shifts on the predictive performance of the trained model. However, the training sets of the clinical datasets exhibited a certain degree of variations caused by patient variance and, in the case of the microscopy datasets, stain variance, thereby introducing a certain degree of \say{natural} color augmentation. All code for model training and evaluation is publicly available in our GitHub repository\footnote{Link will be added upon acceptance of the manuscript.}.  

\begin{table}[t]
\caption{Trainable parameters of baseline and pruned \unet{} architectures.}
\label{tab:unets}
\setlength{\tabcolsep}{3pt}
\centering
\begin{tabular}{p{55pt}p{80pt}p{80pt}}
    \toprule
    & \textbf{ResNet18} &  \textbf{ResNet34}\\
    \midrule
    baseline & \num{14328354} & \num{24436514} \\
    L1-pruned & \num{14309922} & \num{24418082} \\
    L2-pruned & \num{14273058} & \num{24381218} \\
    L3-pruned & \num{14125602} & \num{24233762} \\
    L4-pruned & \num{13535778} & \num{23643938} \\
    \bottomrule
\end{tabular}
\end{table}

\section{Results}
For all models, we first evaluated the segmentation performance using the \ac{iou} computed from the confusion matrix of the segmentation predictions and the ground truth annotation masks accumulated on the complete test set. For the synthetic dataset, we report the cell \ac{iou}, for the \msmf{} dataset the mitotic figure \ac{iou}, for the \mscct{} dataset the tumor \ac{iou}, and for the \mvmr{} dataset the mean \ac{iou} of LV, RV, and MYO. 

\subsection{Synthetic malaria dataset}
\begin{figure*}[t]
\centering
    \begin{subfigure}[b]{\linewidth}
        \includegraphics[height=5cm]{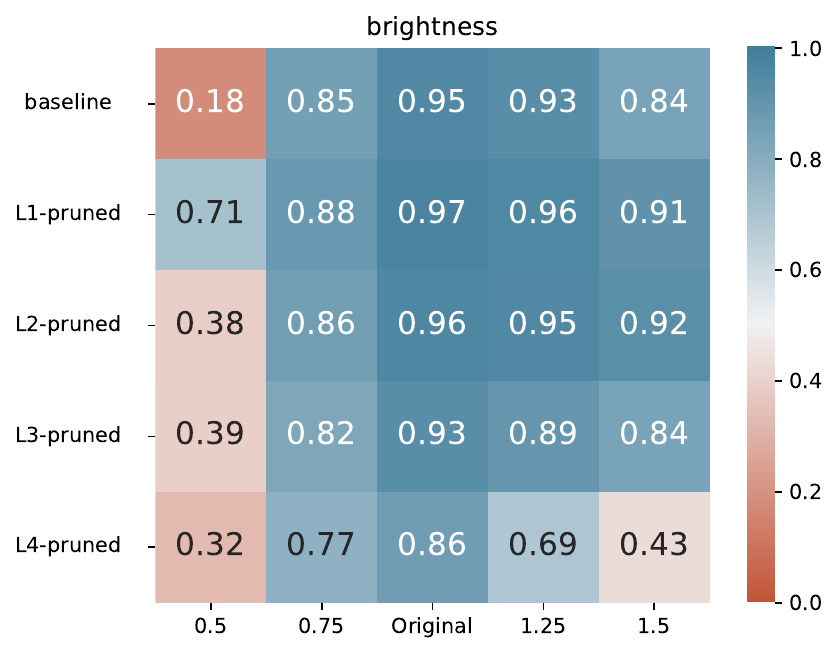} \hspace{-1cm}
        \includegraphics[height=5cm]{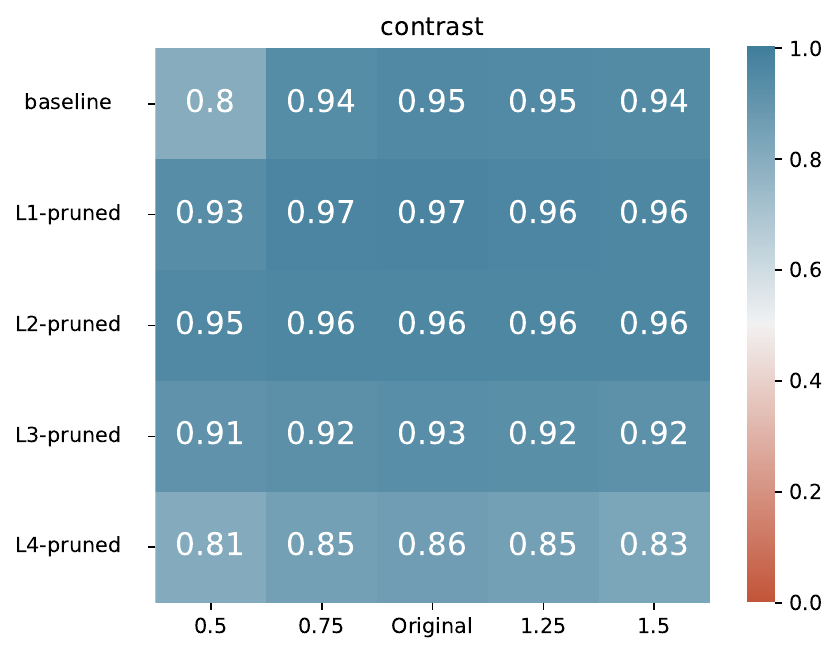} \hspace{-1cm}
        \includegraphics[height=5cm]{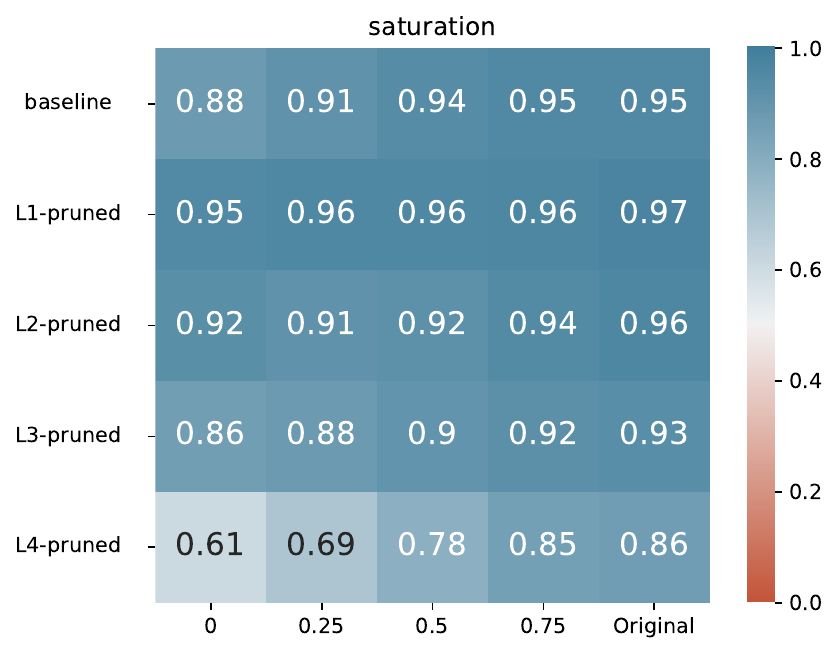}
        \caption{ResNet18 Encoder}
    \end{subfigure}
    \begin{subfigure}[b]{\linewidth}
        \includegraphics[height=5cm]{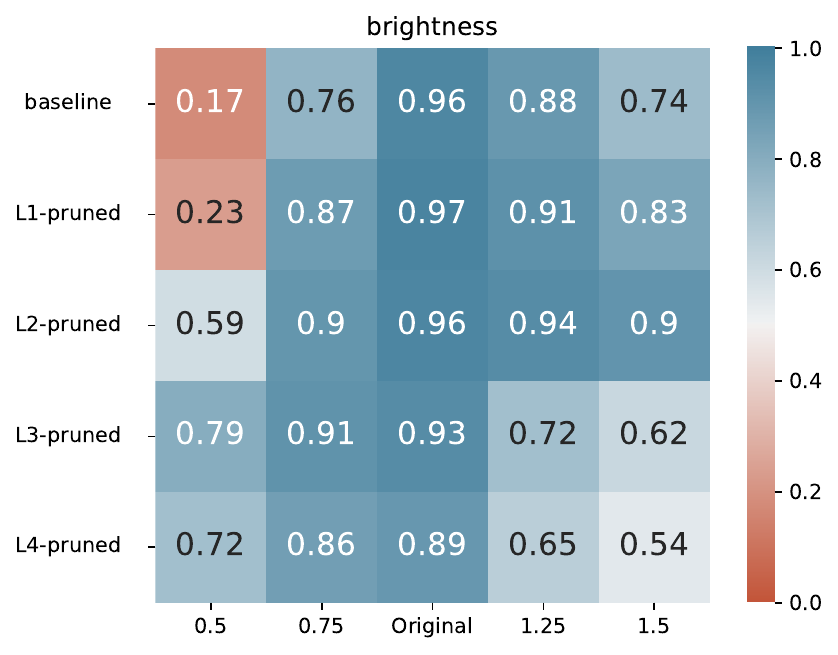} \hspace{-1cm}
        \includegraphics[height=5cm]{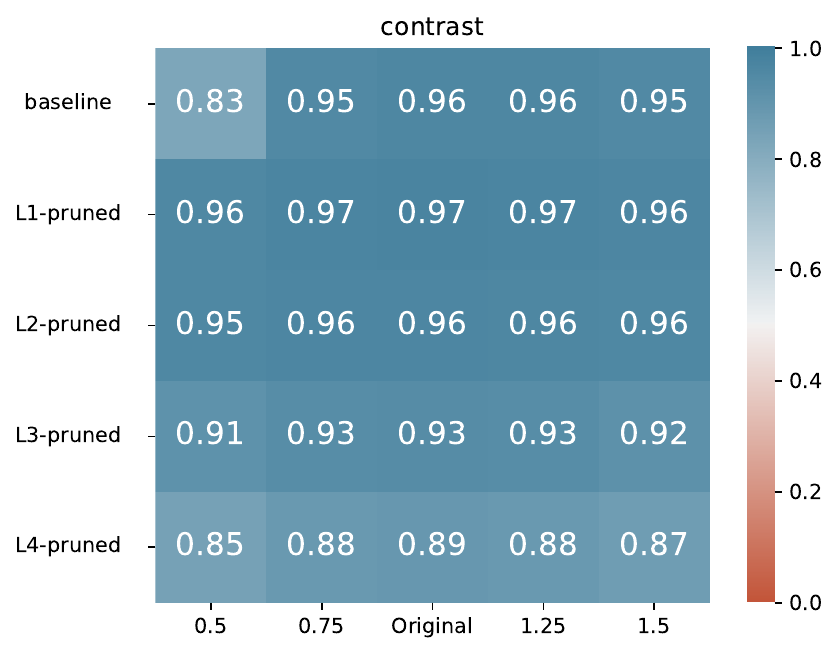} \hspace{-1cm}
        \includegraphics[height=5cm]{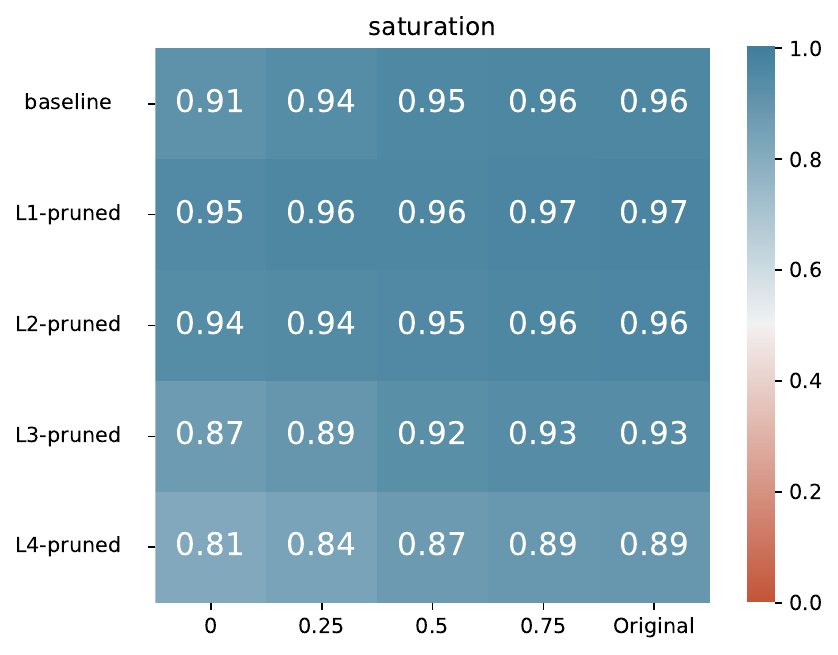}
    \caption{ResNet34 Encoder}
    \end{subfigure}
    \caption{Synthetic malaria dataset. Segmentation performance of trained architectures for different augmentation strengths. The performance was measured as cell \acf{iou} and averaged across all folds of the five-fold cross-validation.} 
    \label{fig:iou_toydata}
\end{figure*}

\Cref{fig:iou_toydata} visualizes the performance of the baseline architecture and the pruned models for different augmentation strengths. The column labeled \emph{original} states the cell \ac{iou} without applying any augmentations. All values were averaged across all folds of the cross-validation. Generally, the models performed best for in-distribution samples and decreased in performance with increasing augmentation strength. The models were most severely affected by changes in brightness, with a decrease in \ac{iou} by up to almost \SI{80}{\percent}. When removing the upper-most skip connection (L1-pruned \unet), the segmentation performance was improved, both cross-domain and, surprisingly, also in-domain. When removing additional skip connections, the pruned models were sometimes less affected by particularly strong augmentations (\eg, left-most column of brightness augmentation for ResNet34 encoder), but generally, the overall performance decreased.      

\begin{figure*}[t]
\centering
\begin{subfigure}[b]{\linewidth}
\includegraphics[width=\linewidth]{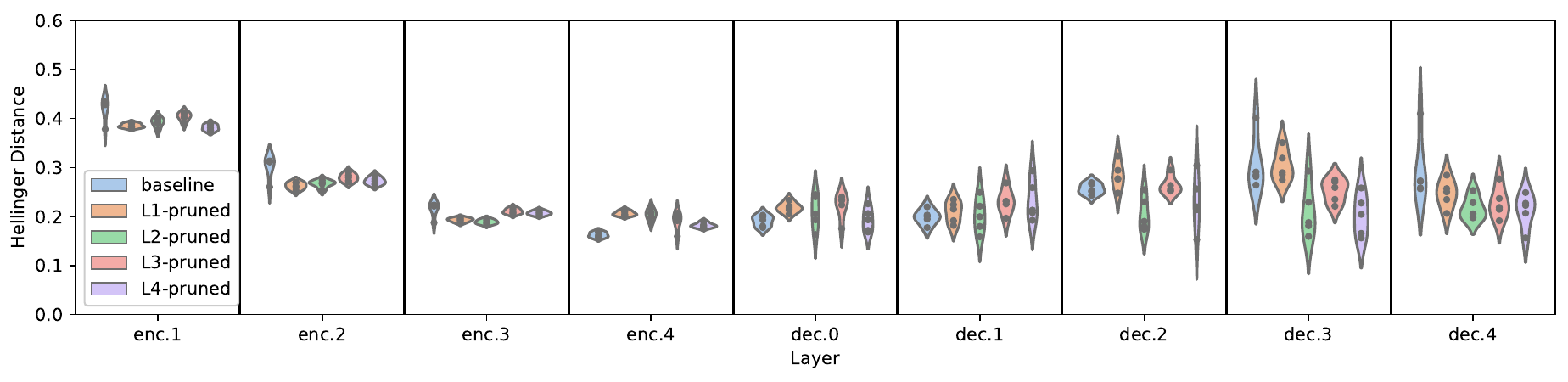}
\caption{ResNet18 Encoder}
\end{subfigure}
\begin{subfigure}[b]{\linewidth}
\includegraphics[width=\linewidth]{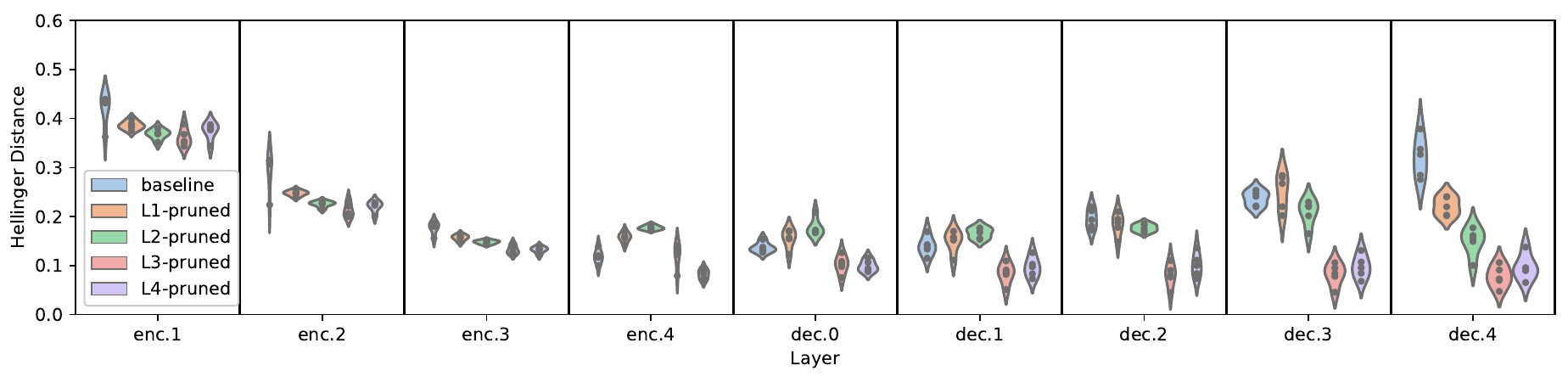}
\caption{ResNet34 Encoder}
\end{subfigure}
\caption{Violin plots of layer-wise Hellinger distance of trained architectures. The distance was assessed using the original test set of the synthetic malaria dataset and an augmented version using a brightness factor of $\lambda_b=0.75$ (corresponding to the second column in the brightness \ac{iou} matrix in \cref{fig:iou_toydata}).}
\label{fig:hellinger_brighness}
\end{figure*}

\Cref{fig:hellinger_brighness} visualizes the layer-wise Hellinger distance of the trained architectures for the original test set and an augmented version. For the visualization, we selected the brightness augmentation with $\lambda_b=0.75$, which corresponds to the second column in the brightness \ac{iou} matrix in \cref{fig:iou_toydata}. The illustration highlights a high domain shift at the first encoder level which was gradually reduced towards the bottleneck. Afterward, the baseline and L1-pruned \unet{} exhibited a slight increase in the Hellinger distance while the other models retained a more or less constant value. The L2-, L3-, and L4-pruned models demonstrated a lower Hellinger distance for the ResNet34 encoder than for the ResNet18 encoder. This corresponds to substantially higher \ac{iou} values of \num{0.90} vs. \num{0.86} (L2), \num{0.91} vs. \num{0.82} (L3), and \num{0.86} vs. \num{0.77} (L4). Generally, the models exhibited the lowest Hellinger distance at the end of the encoder, \ie, after the bottleneck. 

% Hellinger Distance

\begin{figure*}[!ht]
\centering
\begin{subfigure}[t]{0.34\linewidth}
    \includegraphics[height=5cm]{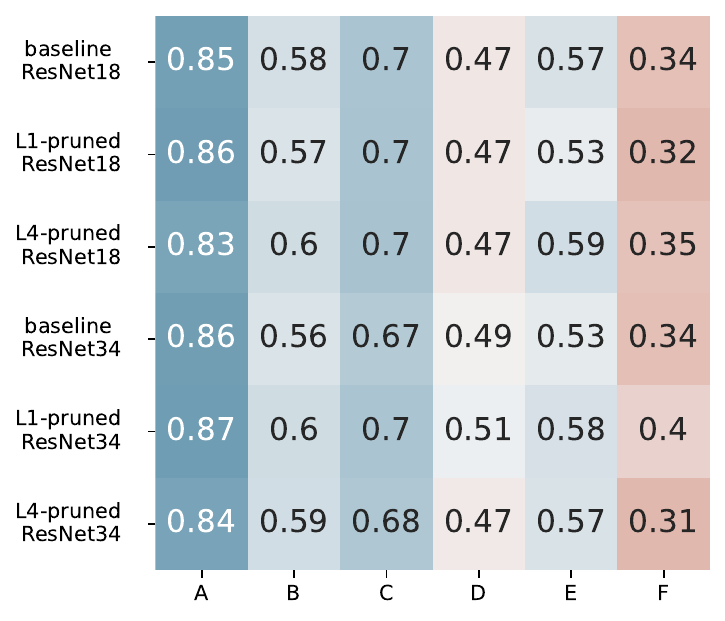} 
\caption{\msmf{} dataset. A:~\mfA{} (Leica), B:~\mfB{} (Hamamatsu), C:~\mfC{} (Hamamatsu), D:~\mfD{} (3DHISTECH), E:~\mfE{} (3DHISTECH), F:~\mfF{} (Philips).}
\label{subfig:mvmf}
\end{subfigure} \hfill
\begin{subfigure}[t]{0.30\linewidth}
\includegraphics[height=5cm]{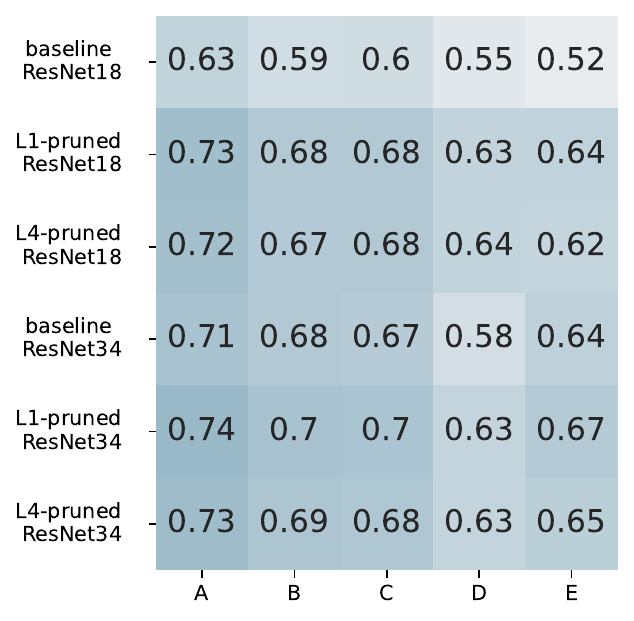} 
\caption{\mscct{} dataset. A:~\sccA{} (Leica), B:~\sccB{} (Hamamatsu), C:~\sccC{} (Hamamatsu), D:~\sccD{} (3DHISTECH), E:~\sccE{} (Leica).}
\label{subfig:mvcct}
\end{subfigure} \hfill
\begin{subfigure}[t]{0.30\linewidth}
\includegraphics[height=5cm]{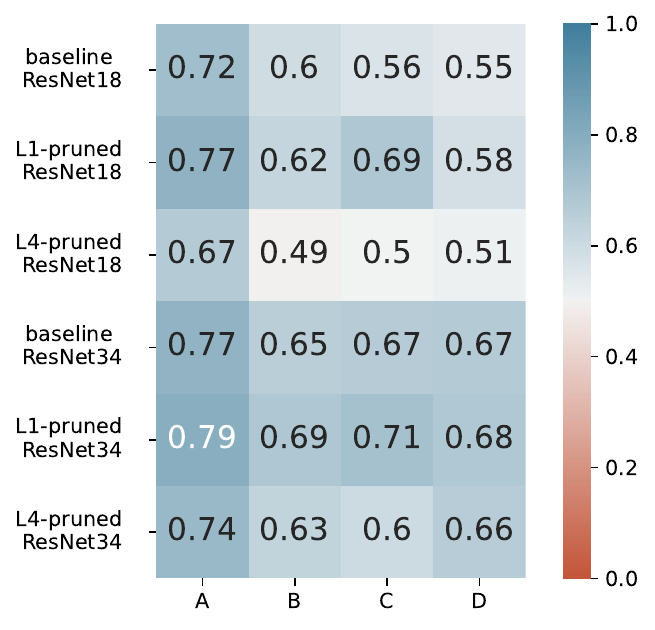}
\caption{\mvmr{} dataset. A:~\mrA{} (Siemens), B:~\mrB{} (Philips), C:~\mrC{} (General Electric), D:~\mrD{} (Canon).}
\label{subfig:mvmr}
\end{subfigure}
\caption{Segmentation performance on medical datasets. The model was trained on domain A and tested across all (WSI / MR) scanners. Values report average segmentation performance across all five folds of the cross-validation.} 
\label{fig:iou_medical}
\end{figure*}

\subsection{\msmf{} dataset}
\Cref{subfig:mvmf} summarizes the mitotic figure \ac{iou} for the tested architectures averaged across the five folds of the cross-validation. Similar to the synthetic dataset, the models performed best for in-domain samples and showed considerable degradation in performance on the \ac{ood} scanners. The model performed worst on the \mfD{} (domain D) and the \mfF{} (domain F) scanning systems. The exemplary patches of these scanners in \cref{fig:mfs} appear brighter and less structured than the remaining patches. It is worth noting that on the synthetic dataset, the \unet{} was most severely affected by changes in brightness, which corresponds to the bright visual appearance of the aforementioned scanners. The L1-pruned architecture again demonstrated a slightly higher in-domain performance. For cross-domain performance, the L1-pruned architecture with a ResNet34 encoder consistently outperformed the baseline \unet{}, especially for strong domain shifts. For the ResNet18 backbone, this is not always the case. Similar to the synthetic dataset, the L4-pruned performed worse than the baseline on in-domain samples, but for some \ac{ood} samples it even performed slightly better than the baseline \unet.

\subsection{\mscct{} dataset}
\Cref{subfig:mvcct} summarizes the tumor \ac{iou} for the tested architectures averaged across the five folds of the cross-validation. On this dataset, the effects of the domain shifts were less severe than on the other histopathology dataset (\msmf), indicated by smaller differences between the in-domain (column A) and \ac{ood} performance (columns B-E). Again, the models achieved the highest tumor \ac{iou} on in-domain samples. On the \mscct{} dataset, the L1-pruned \unet{} demonstrated substantial performance improvements over the baseline \unet{}, both, in- and cross-domain. For the ResNet18 encoder, these improvements were in the range of \SIrange{8}{12}{\percent} and for the ResNet18 encoder in the range of \SIrange{3}{5}{\percent}. On this dataset, the L4-pruned architecture even outperformed the baseline \unet{} but overall achieved a slightly lower performance than the L1-pruned \unet{}.

\subsection{\mvmr{} dataset}
\Cref{subfig:mvmr} summarizes the mean \ac{iou} for the tested architectures averaged across the five folds of the cross-validation. The models performed best on in-domain samples and, for the baseline \unet{}, performance decreases by up to \SI{17}{\percent} could be observed for \ac{ood} samples (columns B-D). The L1-pruned \unet{} again consistently outperformed the baseline, whereas the L4-pruned \unet{} resulted in a lower performance.

\section{Discussion and outlook}
The experiments on the synthetic malaria dataset demonstrated substantial performance drops when applying the trained segmentation model on \ac{ood} test data. These were most severe for changes in brightness but could also be observed for changes in contrast and saturation. The layer-wise analysis of the Hellinger distance highlighted the model-inherent domain shift in the feature encoding of the original dataset and the augmented version. These differences were more pronounced at earlier layers of the encoder, which is in line with observations from previous studies. When removing the uppermost skip connection (L1-pruned), the model consistently achieved higher \acp{iou} not only cross-domain but surprisingly also for in-domain samples. This in-domain performance increase could be observed across all datasets and was even more pronounced on the clinical datasets than on the synthetic dataset. Due to a certain degree of patient and stain variance, the clinical datasets might exhibit a hidden domain shift to which the L1-pruned architecture seems to be more robust. This indicates that removing skip connections may contribute substantially to the robustness of a segmentation model and potential skip connection pruning should be considered during model hyperparameter tuning.

The observations made in this work have implications for related research areas in the realm of deep learning. Many recent works have studied \ac{ssl} to leverage large unlabeled datasets for pre-training \acp{cnn} to enable the model to learn generic visual concepts about the data and potentially make \acp{cnn} more robust to domain shifts~\cite{li2021domain}. Most of these works, however, have focused on classification tasks and recent studies have shown that these performance benefits do not necessarily translate to dense prediction tasks~\cite{ericsson2021well, wilm2023mind}, such as image segmentation. Yang  \etal~\cite{yang2021instance} hypothesized that the architectural changes that have to be undergone when re-purposing a pre-trained encoder for dense prediction tasks limit their transferability. The risk of re-introducing domain-specific features across skip connections, as demonstrated here, supports this theory and these implications for \ac{ssl} could be grounds for future work.             

\bibliographystyle{IEEEtran}
\bibliography{bibliography}

\begin{acronym}
\acro{catch}[CATCH]{CAnine cuTaneous Cancer Histology}
\acro{cmc}[CMC]{canine mammary carcinoma}
\acro{gdv}[GDV]{Generalized Discrimination Value}
\acro{kl}[KL]{Kullback-Leibler}
\acro{iou}[IoU]{intersection over union}
\acro{miou}[mIoU]{mean intersection over union}
\acro{mr}[MR]{magnetic resonance}
\acro{cnn}[CNN]{neural network}
\acro{lut}[LUT]{lookup table}
\acro{ood}[OOD]{out-of-distribution}
\acro{ssl}[SSL]{self-supervised learning}
\acro{scc}[SCC]{squamous cell carcinoma}
\acro{wsi}[WSI]{whole slide image}
\end{acronym}

\end{document}